\documentclass[
twocolumn,
preprintnumbers,amsmath,amssymb,aps,prd,nofootinbib]{revtex4}

\usepackage{graphicx}
\usepackage{dcolumn}
\usepackage{bm}
\newcommand\be{\begin{equation}}
\newcommand\ee{\end{equation}}
\newcommand\bea{\begin{eqnarray}}
\newcommand\eea{\end{eqnarray}}

\begin{document}

\title{Super-horizon cosmic string correlations}

 \author{Arttu Rajantie}%
 \email{a.rajantie@imperial.ac.uk}
\affiliation{%
Department of Physics,
Imperial College London,
Prince Consort Road,
London SW7 2AZ,
UK}%

\date{28 January 2009}

\begin{abstract}
When gauged cosmic strings form in a symmetry-breaking phase transition, the gauge field configuration at the time becomes imprinted in the spatial string distribution by the flux trapping mechanism. Causality and flux conservation suggest that that quantum and thermal gauge field fluctuations give rise to long-range superhorizon correlations in the string network. Classical field theory simulations in the Abelian Higgs model confirm this finding. In contrast, the Kibble-Zurek mechanism which most cosmic string studies are based on, only gives rise to short-distance, subhorizon correlations. These results may have implications for cosmology, and it may also be possible to test them in superconductor experiments.
\end{abstract}

\preprint{Imperial/TP/08/AR/01}

\maketitle

\section{Introduction}
\label{sec:intro}

Cosmic strings~\cite{Kibble:1976sj,Hindmarsh:1994re,Sakellariadou:2006qs} are hypothetical string-like objects, which are stable and extremely massive and whose existence is predicted by many field theory~\cite{Jeannerot:2003qv} and string theory models~\cite{Copeland:2003bj} of particle physics. If they exist in nature, they would have generally been formed in the early universe in symmetry breaking phase transitions~\cite{Kibble:1976sj}, or in brane collisions~\cite{Dvali:2003zj} in brane world models. 

Cosmic strings are cosmologically interesting, because their energy density tracks the overall energy density of the universe~\cite{Kibble:1984hp}. Most other stable objects produced in the early universe would eventually dominate the energy density, in clear contradiction with observations, but strings always contain a small but non-negligible fraction of the total energy. 

Because of this scaling behaviour, cosmic strings would introduce nearly scale invariant density perturbations, and this made them at first a strong candidate for explaining the origin of primordial density perturbations. Even the amplitude of the perturbations was consistent with cosmic strings in typical Grand Unified Theories. However, acoustic peaks observed in the cosmic microwave background~\cite{Netterfield:2001yq} and large scale structure~\cite{Eisenstein:2005su} showed that perturbations existed on super-horizon scales.  This could not be explained by the cosmic string model and provided the decisive piece of evidence in favour of inflation. 

Because cosmic strings are a fairly generic feature of unified theories~\cite{Jeannerot:2003qv}, they are still being searched for in many different ways. If they exist, they would give a potentially observable contribution to the temperature anisotropies of the cosmic microwave background radiation~\cite{Kaiser:1984iv}. They would also radiate gravitational waves and lens light from distant objects. These effects have been used to constrain cosmic string models~\cite{Bevis:2006mj,Jenet:2006sv,Christiansen:2008vi}.

The standard assumption in cosmic string studies is that the strings are formed by the Kibble-Zurek mechanism~\cite{Kibble:1976sj,Zurek:1985qw,Rajantie:2001ps}. In that case, the network would be uncorrelated at long distances, and the scaling behaviour follows naturally. In numerical simulations of cosmic string evolution, this assumption is introduced in the form of random (Vachaspati-Vilenkin) initial conditions for the scalar field~\cite{Vachaspati:1984dz}.

The Kibble-Zurek picture is believed to be an accurate description of the formation of global cosmic strings~\cite{Rajantie:2001ps}. However, global strings have logarithmically confining long-range interactions, which makes their evolution and cosmological effects more complicated. The usual picture of line-like strings with short-range interactions, which most cosmological studies are based on, applies only to gauged strings. Furthermore, cosmic strings predicted by realistic field theory and string theory models tend to be gauged rather than global~\cite{Jeannerot:2003qv}. 

The formation of gauged strings takes place differently because two fields, the scalar and the gauge field, are involved. Strings are formed not only by the Kibble-Zurek mechanism but also when gauge field fluctuations get trapped during the transition~\cite{Hindmarsh:2000kd}. The predictions of this flux trapping mechanism have been confirmed in numerical simulations~\cite{Stephens:2001fv,Rajantie:2001ps,BlancoPillado:2007se}.

This paper investigates the flux trapping mechanism and, in particular, the effects it has on the spatial distribution of strings. It is shown that any pre-existing features in the gauge field configuration become imprinted on the cosmic string network. If these features extend to super-horizon scales, which is quite possible in inflationary cosmology, they can influence the later evolution of the network and its observable consequences. Furthermore, these arguments suggest that even if the gauge field is initially in its vacuum state, the quantum vacuum fluctuations themselves give rise to super-horizon long-range correlations in the string network.

\section{Symmetry Breaking and Cosmic Strings}
\label{sec:ssb}
The simplest model of cosmic strings is the Abelian Higgs model in which strings correspond to Nielsen-Olesen vortex lines~\cite{Nielsen:1973cs}. This paper will focus on this model, but the main conclusions are based on general physical principles and should be applicable much more broadly. They should even apply to at least certain cases of cosmic superstring formation in brane inflation models~\cite{Copeland:2003bj,Dvali:2003zj}.

The Abelian Higgs model consists of a complex scalar $\phi$ and a
gauge field $A_\mu$, with the Lagrangian
\be
\label{equ:lag}
{\cal L}_{\rm gauge}=-\frac{1}{4}F_{\mu\nu}F^{\mu\nu}+(D_\mu\phi)^*D^\mu\phi-V(\phi),
\ee
where $D_\mu=\partial_\mu+ieA_\mu$ is the covariant derivative, $F_{\mu\nu}=\partial_\mu A_\nu-\partial_\nu A_\mu$ is the field strength tensor, and $e$ is the gauge coupling constant. We assume that the potential $V(\phi)$ is of the renormalisable form
\be
\label{equ:pot}
V(\phi)=m^2\phi^*\phi+\lambda(\phi^*\phi)^2.
\ee
This Lagrangian is invariant under local {\em gauge} transformations,
\be
\label{equ:gaugetransform}
\phi(x)\rightarrow e^{i\Delta(x)}\phi(x),\quad A_\mu(x)\rightarrow A_\mu(x)-\frac{1}{e}\partial_\mu\Delta(x),
\ee
where the rotation angle $\Delta$ can depend on the spacetime position.
Physically, the field $\phi$ describes charged scalar particles with long-range interactions mediated by the gauge field. In the cosmological setting, the U(1) gauge group is not that of electrodynamics, although it has the same qualitative properties. In this paper, the components of the field strength tensor $F_{\mu\nu}$ will nevertheless be referred to as ``electric'' and ``magnetic'' fields.

In the limit $e\rightarrow 0$, the gauge field decouples, and the Lagrangian reduces to
\be
\label{equ:globallag}
{\cal L}_{\rm global}=\partial_\mu\phi^*\partial^\mu\phi-V(\phi).
\ee
This Lagrangian is only invariant under {\em global} phase rotations $\phi\rightarrow e^{i\Delta}\phi$,
where $\Delta$ is constant over the whole spacetime. Physically, the quanta of $\phi$ are therefore ``electrically'' neutral.

In both cases, when $m^2$ is positive, the vacuum state, given by the minimum of the potential, corresponds to $\phi=0$. For negative $m^2$, there is a set of degenerate vacua
\be
\label{equ:vacua}
\phi=ve^{i\theta},\quad v^2\equiv -\frac{m^2}{2\lambda},
\ee
parameterized by the complex phase $-\pi< \theta\le\pi$.
Because these vacua are related by a symmetry, they are all identical. Once the system has chosen a particular vacuum, its state is no longer invariant under symmetry transformations, and the symmetry is said to be spontaneously broken.
In the gauge theory, the covariant derivative term gives the gauge field a mass $m_\gamma=ev$.

In both models there are line-like stable topological defect solutions in the broken phase. The solution has the form (in cylindrical polar coordinates $r$, $\varphi$, $z$)
\be
\label{equ:vortexphi}
\phi(r,\varphi,z)=vf(r)e^{iN_w\varphi},
\ee
where $N_w$ is an integer winding number, and $f(r)$ is a function that has to be determined numerically and satisfies the boundary conditions
\be
f(r)\rightarrow
	\begin{cases}
	0,&\mbox{as $r\rightarrow 0$,}\cr
	1,&\mbox{as $r\rightarrow \infty$.}
	\end{cases}
\ee

In the gauge theory, the gauge field in the appropriate gauge is
\be
\label{equ:vortexA}
\vec{A}=\frac{N_w}{er}a(r)\hat\varphi,
\ee
where $a(r)$ is a function that has to be determined numerically and satisfies the same boundary conditions as $f(r)$. Because the magnetic flux is given by the contour integral
\be
\label{equ:fluxdef}
\Phi=\oint d\vec{r}\cdot\vec{A}=\frac{2\pi}{e}N_w,
\ee
the strings carry a quantized magnetic flux. In fact they are fully analogous to Abrikosov flux tubes in superconductors~\cite{Abrikosov:1956sx}. Outside the vortices, the magnetic field decays exponentially, as a manifestation of the Meissner effect.

In the gauge theory (\ref{equ:lag}), the energy of the string is localised exponentially on the string, and these gauged strings behave like idealised Nambu-Goto strings~\cite{Allen:1990tv,Hindmarsh:1994re}. In contrast, energy of a string is not localised in the global theory (\ref{equ:globallag}). The strings have a confining logarithmic interaction, and therefore these global strings evolve differently, and much less is known about their cosmological effects.

If a network of cosmic strings was formed in the early universe, its later evolution will have produced potentially observable effects such as gravitational wave emission and temperature anisotropies in the cosmic microwave background (CMB) radiation. All the observable effects are gravitational and depend therefore on the dimensionless combination $G\mu$, where $G$ is Newton's constant and $\mu$ is the string tension, defined as the energy per unit length. 

The evolution of the string network is a complicated non-linear problem, and even with large-scale numerical simulations one can follow it only for a relatively short time~\cite{Bevis:2006mj}. To make predictions, one usually assumes that the evolution is such that the network scales with the horizon size~\cite{Bennett:1989ak,Copeland:1991kz}. As will be discussed in more detail in Section~\ref{sec:cosmo}, this scaling behaviour follows naturally if there is sufficient energy loss and if the string network is uncorrelated at long distances.

The scaling behaviour means that at any time, the network looks statistically the same in units of $1/H$. With this assumption, the total length of the string network inside one Hubble volume scales as $H^{-1}$, indicating that the total energy is $\sim \mu/H$. Dividing by the volume $\sim H^{-3}$, we find the energy density,
\be
\rho\sim \mu H^2=\frac{8\pi}{3}G\mu\rho_c,
\ee
where $\rho_c$ is the critical energy density. Therefore as long as the scaling assumption is valid, the strings contribute a constant fraction of the total energy of the universe, and will never dominate or be negligible. This is why they are cosmologically more interesting than other topological defects, i.e., domain walls or monopoles, which tend to dominate the energy density and overclose the universe.

As a consequence of the scaling, the perturbations generated by the string network are scale invariant as required by the observations. However, in the absence of super-horizon correlations, they would not have the observed acoustic peaks~\cite{Allen:1996wi}. The contribution to the density perturbations from cosmic string must therefore be subdominant. This sets an upper limit for the string tension $G\mu\lesssim 10^{-5}$.

\section{Local String Density}
\label{sec:density}
To discuss correlations in the string network, Liu and Mazenko~\cite{LiuMazenko} defined 
the local string density $\vec{\rho}$ as
\be
\label{equ:rhodef0}
\vec\rho(\vec{x})=\sum_\alpha\int ds \frac{d\vec{x}_\alpha}{ds}\delta\left(\vec{x}-\vec{x}_\alpha(s)\right),
\ee
where $\alpha$ labels the strings and $s$ is a coordinate along the string. In the Abelian Higgs model, this can be expressed in terms of the fields as
\be
\label{equ:rhodef}
\vec{\rho}=\vec{\nabla}\times\vec{q},
\ee
where $\vec{q}$ is defined as
\be
\label{equ:qdef}
\vec{q}=
\frac{1}{2\pi}\left({\rm Im}\,\hat\phi^*\vec{D}\hat\phi - e\vec{A}\right)
=
\frac{1}{2\pi}\vec{\nabla}\theta.
\ee
Here $\theta=\arg\phi$ is the complex phase of the scalar field $\phi$, and $\hat\phi=\phi/|\phi|=\exp(i\theta)$.
The local string density is gauge invariant and sourceless
\be
\label{equ:sourceless}
\vec{\nabla}\cdot\vec{\rho}=0.
\ee
It vanishes outside the strings and 
has a delta-function peak tangential to the string at the centre of the string.

The two-point string correlation function $G_{ij}(\vec{x})$ is defined as
\be
\label{equ:corrdef}
G_{ij}(\vec{x})=\langle \rho_i(0) \rho_j(\vec{x})\rangle.
\ee
It is also often useful to consider the Fourier transform of this, 
\be
\label{equ:FFTcorr}
G_{ij}(\vec{k})=\int d^3x e^{i\vec{k}\cdot\vec{x}}
G_{ij}(\vec{x}),
\ee
which gives the momentum space correlation function
\be
\langle \rho_i(\vec{k}) \rho_j(\vec{k}')\rangle =
G_{ij}(\vec{k})(2\pi)^3\delta\left(\vec{k}+\vec{k}'\right).
\ee
Because of rotation invariance and the sourcelessness condition (\ref{equ:sourceless}), 
this can be written as
\be
\label{equ:corrtensor}
G_{ij}(\vec{k})=\left(\delta_{ij}-\frac{k_ik_j}{k^2}\right)G(k),
\ee
where $G(k)$ is a scalar and can be obtained from the trace of $G_{ij}$,
\be
\label{equ:corrtrace}
G(k)=\frac{1}{2}G_{ii}(\vec{k}).
\ee

Note that the correlator $G$ is not the same as the correlator of the energy-momentum tensor~\cite{Turok:1996ud,Vincent:1996qr,Bevis:2006mj}, which is more commonly discussed and which seeds density perturbations. Long-range correlations in one quantity do not imply long-range correlations in the other. However, as will be discussed in Section~\ref{sec:cosmo}, long-range correlations in $\vec{\rho}$ will affect the time evolution of the network and thereby also indirectly the density perturbations and other observable signatures.

\section{String Formation}
\label{sec:formation}

Cosmic strings were formed in the early universe if a U(1) symmetry became spontaneously broken in a phase transition. In the original scenario~\cite{Kibble:1976sj}, the transition took place at a high temperature. To first approximation, the effects of the temperature are given by an additive temperature-dependent contribution to the mass parameter $m^2$, so that the effective mass parameter $m^2(T)$ is
\be
m^2(T)\approx m^2+\left(\frac{1}{4}e^2+\frac{1}{3}\lambda\right) T^2,
\ee
where $m^2<0$. At high temperatures,
\be
T>T_c\approx\sqrt{\frac{-12m^2}{3e^2+4\lambda}},
\ee 
the effective mass parameter is positive and the symmetry is restored. When the universe cooled down and the temperature decreased below the critical value, $m^2(T)$ became negative and the symmetry was broken.

In inflationary models, string formation can take place at the end of inflation, 
and in this case the temperature is generally extremely low. In hybrid inflation~\cite{Linde:1993cn}, the field $\phi$ is coupled to the inflaton field $\sigma$, so that the effective mass is
\be
m^2(\sigma)=m^2+g^2\sigma^2,
\ee
where $g$ is the coupling constant for the $\phi-\sigma$ interaction.
When the inflaton rolls below the critical value $\sigma_c=(-m^2/g^2)^{1/2}$, the symmetry gets broken. This transition not only forms strings but also triggers the end of inflation.

Brane collisions in brane world and string theory models are effectively very similar to the phase transition in hybrid inflation~\cite{Dvali:2003zj}. In general, the role of the inflaton is played by the distance between the branes, and when the branes collide, they annihilate through tachyon condensation, which is effectively a symmetry-breaking phase transition.

Whatever the microphysics that causes the symmetry breaking, it generally forms cosmic strings. There are two distinct physical mechanisms that are responsible for this: the Kibble-Zurek mechanism~\cite{Kibble:1976sj,Zurek:1985qw} and flux trapping~\cite{Hindmarsh:2000kd,Rajantie:2001ps}.

\subsection{Kibble-Zurek Mechanism}
\label{sec:KZ}

The Kibble-Zurek mechanism involves only the scalar field $\phi$, and therefore it operates in both global and gauge theories. In the gauge theory, it is only part of the dynamics and becomes dominant if the gauge field can be ignored.

When the symmetry is broken, the scalar field has to choose one of the possible vacua (\ref{equ:vacua}), corresponding to one particular complex phase $\theta$. All values of $\theta$ are equally likely because all the vacua are identical. Kibble's original observation~\cite{Kibble:1976sj} was that in cosmology, any two points that are separated by more than the horizon size are causally disconnected, and they will have to make this choice independently of each other. After the transition, the universe will therefore consist of finite regions, which cannot be larger than the horizon, in each of which the choice of the vacuum is random, and between which it is uncorrelated.

If one follows the complex phase around a closed curve that passes through at least three of these regions, there is a certain non-zero probability that it changes by a multiple of $2\pi$. In this case, continuity implies that the field has to vanish somewhere inside the curve, and this corresponds to the core of the string. According to this argument there should be roughly one string per correlated region, whose size is limited by the horizon size.

Zurek~\cite{Zurek:1985qw} showed later that the size of the correlated regions is generally determined not by the cosmological horizon size, but by the critical slowing down during the transition. The correlation length $\xi$ grows as the critical temperature is approached, and in an adiabatic transition it would diverge at the transition point. However, the phase transition can never be adiabatic. The dynamics of the system slows down when the transition is approached, and if the transition takes place in a finite time, the system cannot stay in equilibrium. Instead of actually diverging, the correlation length remains finite, freezing out at some finite value $\hat\xi$, which determines the size of the correlated regions. The ultimately upper bound for this is given by causality, since the correlation length cannot grow faster than the speed of light. From this argument Zurek was able to derive a scaling law for the number of defects as function of the cooling rate, which has been confirmed in experiments~\cite{Rajantie:2001ps}.

Since the complex phase $\theta$ is only correlated at distances less than $\hat\xi$, Eqs.~(\ref{equ:rhodef}) and (\ref{equ:qdef}) imply that the same must be true for the string density $\vec{\rho}$. Therefore the correlator must have a finite range. For instance, Liu and Mazenko find a Gaussian correlator~\cite{LiuMazenko}
\be
G_{ij}(\vec{x})\sim e^{-\vec{x}^2/2\hat\xi^2}.
\ee
Whatever the precise form of the correlator, there are no long-range correlations in a string network formed by the Kibble-Zurek mechanism. 

Furthermore, the integral of the correlator over the whole space vanishes. It can be expressed as a surface integral over its boundary,
\begin{eqnarray}
\label{equ:KibbleGint}
\int d^3x G_{ij}(\vec{x})&=&
\left\langle
\rho_i(0)\int d^3x \rho_j(\vec{x})\right\rangle\nonumber\\
&=&\epsilon_{jkl}\int d^2S_k \left\langle \rho_i(0)q_l(\vec{x})\right\rangle.
\end{eqnarray}
In this integral all points $\vec{x}$ are on the boundary, and because of the finite correlation length, 
the correlator $\langle \rho_i(0)q_l(\vec{x})\rangle$ and therefore the right-hand side of Eq.~(\ref{equ:KibbleGint}) vanishes, i.e.,
\be
\label{equ:KZcons}
\int d^3x G_{ij}(\vec{x})=0.
\ee
This indicates a negative correlation between strings: The winding number of a string is screened by the windings of nearby strings.

Because of the finite range, the momentum space correlator $G(k)$ is analytic in $k^2$ and can be Taylor expanded,
\begin{eqnarray}
\label{equ:KZtaylor}
G(k)&=&\frac{2\pi}{k}\int_0^\infty dr\,r\sin(kr) G_{ii}(r)\nonumber\\
&=&
4\pi\int_0^\infty dr\,r^2
\left(1-\frac{k^2r^2}{6}\right)G(r)+O(k^4)
\nonumber\\
&\sim& \hat\xi k^2+O(k^4).
\end{eqnarray}
The constant term cancels because of Eq.~(\ref{equ:KZcons}), and the $k^2$ term has to have this form because $\hat\xi$ is the only available dimensionful scale.

\subsection{Flux Trapping}
\label{sec:flux}

In the gauge theory, strings can also form by flux trapping~\cite{Hindmarsh:2000kd,Rajantie:2001ps}. Magnetic fields cannot exist in the broken phase vacuum, and if a non-zero field is present at the time of the transition, it must therefore become confined in strings. The simplest example is a transition in a uniform magnetic field, which is a common setup for superconductor experiments. The Meissner effect tries to expel the field but inside the superconductor it is unable to do that, and instead it forms a lattice of Abrikosov vortices~\cite{Abrikosov:1956sx}, where the average density of strings is directly related to the field strength,
\be
\label{equ:rhoB}
\vec\rho=-\frac{e}{2\pi}\vec{B}.
\ee

In the early universe there was no external field, but there were thermal and quantum fluctuations. Because the photon is massless in the symmetric phase, these fluctuations can have arbitrarily long wavelengths, or equivalently, arbitrarily low wave numbers $\vec{k}$. When the system enters the broken phase, magnetic field modes with wave number less than the photon mass, $|\vec{k}|<m_\gamma$, are prohibited. The system tries to suppress these modes to minimize its energy, but the magnetic flux is conserved as a consequence of the Maxwell equation
\be
\frac{\partial \vec{B}}{\partial t}+\vec{\nabla}\times\vec{E}=0.
\ee
This means that the net flux through a surface can only change by a boundary contribution, when magnetic field lines move across the boundary,
\be
\label{equ:fluxcons}
\frac{\partial \Phi}{\partial t}=-\oint d\vec{r}\cdot\vec{E}.
\ee
Because of this, modes with longer wavelengths take more time to decay. Essentially a mode with wavenumber $\vec{k}$ cannot decay in shorter time than what it takes for light to travel one wavelength $\lambda=1/|\vec{k}|$.

A long-wavelength magnetic field mode that has not had time to decay before the phase transition is completed remains in the system even though it is now in the broken phase. As with a uniform field discussed above, this is only possible if it is confined inside strings. The string density (averaged over suitable volume) is then locally related to the magnetic field strength by the same equation (\ref{equ:rhoB}).

This means that long-wavelength fluctuations break up into strings in the same way as a uniform field. This is natural because a fluctuation with long enough wavelength (for instance longer than the cosmological horizon size) is locally indistinguishable from a uniform field.

On the other hand, short-wavelength fluctuations behave adiabatically and decay, which means that there must be a critical wavenumber $\hat{k}$ separating the two different behaviours: Modes with $|\vec{k}|\gtrsim\hat{k}$ decay but modes with $|\vec{k}|\lesssim\hat{k}$ become trapped into strings. The critical wavenumber depends on the dynamics of the system and is generally lower for slower transitions. In principle, its precise value can be calculated for any particular scenario, but that will not be necessary for the conclusions drawn this paper.

If there was any feature in the initial magnetic field configuration that was larger than the critical length scale $1/\hat{k}$, this feature would survive the transition and be imprinted in the cosmic string network. For instance, a region with a strong uniform magnetic field would create a bunch of nearly parallel cosmic strings, which would roughly follow the initial magnetic field lines. Even in the absence of such specific features, which are unlikely in typical cosmological models, thermal and quantum fluctuations lead to non-trivial effects, and as discussed in the next section.

\section{Correlations from Flux Trapping}
\label{sec:FTcorr}

Because the magnetic field is sourceless, its momentum space two-point correlation function has the same tensor structure as Eq.~(\ref{equ:corrtensor}),
\be
\label{equ:GBdef}
\langle B_i(\vec{k}) B_j(\vec{k}')\rangle =
G^B_{ij}(\vec{k})(2\pi)^3\delta\left(\vec{k}+\vec{k}'\right),
\ee
where
\be
\label{equ:GBtensor}
G^B_{ij}(\vec{k})=
\left(\delta_{ij}-\frac{k_ik_j}{k^2}\right)G^B(k),
\ee
and $G^B(k)$ is a scalar function of $k=|\vec{k}|$.
The coordinate space correlator $G^B_{ij}(\vec{x})$ is given by the Fourier transform
\be
\label{equ:GBx}
G^B_{ij}(\vec{x})=\int \frac{d^3k}{(2\pi)^3}e^{i\vec{k}\cdot\vec{x}}G^B_{ij}(\vec{k}).
\ee

The conservation of magnetic flux~(\ref{equ:fluxcons})  means that the correlator can only change by a spatial derivative. In Fourier space, the change in $G^B$ is at least quadratic in $k$,
\be
G^B_{ij}(t;\vec{k})=G^B_{ij}(0;\vec{k})+(k^2\delta_{ij}-k_ik_j)\Delta(k).
\ee

Once the system enters the broken phase, the long-wavelength magnetic field modes freeze out as discussed in Section~\ref{sec:flux}. The final correlation function $G^B_{\rm final}$  will therefore interpolate between the initial on $G^B_{\rm ini}$ at low wave number $k\lesssim\hat k$ and the equilibrium correlator $G^B_{\rm eq}$ at high wave number $k\gtrsim\hat k$. 

The precise form of the correlator is not important, but one obtains a useful approximation by assuming that the long-distance dynamics are diffusive. Then each Fourier mode decays as
\be
\vec{B}(t,\vec{k})=\vec{B}(0,\vec{k})e^{-k^2t/4\pi\sigma},
\ee
where $\sigma$ is the electric conductivity. At some time $\hat{t}$ the dynamics become non-linear when strings form, and one can identify
\be
\hat{k}=\sqrt\frac{2\pi\sigma}{t},
\ee
so that the final correlator is
\be
\label{equ:diffBcorr}
G^B_{\rm final}(k)=G^B_{\rm ini}(k)e^{-k^2/2\hat{k}^2}.
\ee

When the dynamics become non-linear, these magnetic field fluctuations are turned into cosmic strings. If one ignores the contribution from the Kibble-Zurek mechanism, so that flux trapping is the only source of strings, Eq.~(\ref{equ:rhoB}) implies that the string density correlator is simply given by
\be
\label{equ:stringcorr}
G(k)=\frac{e^2}{4\pi^2}G^B_{\rm final}(k)
    =\frac{e^2}{4\pi^2}G^B_{\rm ini}(k)e^{-k^2/2\hat{k}^2}.
\ee

To calculate the correlations of strings formed by flux trapping, one needs to know the spectrum of initial fluctuations $G^B_{\rm ini}$, which depends on the scenario. The two simplest and most relevant cases, when the field in initially in vacuum or in thermal equilibrium, are discussed below.

\subsection{Vacuum Fluctuations}
\label{sec:vacuum}

The simplest situation is when the gauge field is in its vacuum state at the time of string formation. This is the case after a long period of inflation, which would have wiped out any excitations. Because the gauge field is conformally invariant, no long-range fluctuations are generated during inflation~\cite{Turner:1987bw}. Therefore the expansion of the universe simply redshifts and dilutes any existing gauge field configuration, taking it to its vacuum state.

In this case, the symmetry breaking phase transition can be thought of as a non-adiabatic change of the vacuum state of the system. Normally this leads to pair production of particles, which is often be described using Bogoliubov transformations within linear theory with a time-dependent Hamiltonian. 
Our case is physically very similar but the produced excitations are cosmic strings, and because they are non-linear objects, the linear approximation is not valid. In contrast, the physical picture presented in Section~\ref{sec:FTcorr}, which is based on causality and conservation laws, should still be applicable.

It is important to note that even if the universe reheated between the end of inflation and the phase transition, the cosmologically relevant modes would still remain at zero temperature, because they are outside the horizon and can therefore not be affected by local reheating dynamics. Therefore the calculations in this Section should also apply to that situations. Furthermore, these calculations are not restricted to cosmological scenarios but apply generally to any system at zero temperature.

In the symmetric phase, the vacuum two-point correlation function of the magnetic field can be calculated from the photon Feynman propagator,
\be
D_F^{\mu\nu}(k)
=-\frac{i\eta^{\mu\nu}}{k^2+i\epsilon},
\ee
where we have adopted the Feynman gauge.
This is unaffected by the expansion of the universe because the gauge field is conformally invariant.
The equal-time two-point correlator of the gauge field is
\bea
\langle A_i(\vec{k})A_j(\vec{q})\rangle
&=&(2\pi)^3\delta(\vec{k}+\vec{q})
\int_{-\infty}^\infty\frac{dk_0}{2\pi}D_F^{ij}(k_0,\vec{k})
\nonumber\\
&=&\frac{\delta_{ij}}{2k}(2\pi)^3\delta(\vec{k}+\vec{q}).
\eea
The magnetic field correlator is obtained by calculating the curl of the gauge field, 
$B_i(\vec{k})
=i\epsilon_{ijk} k_jA_k(\vec{k}),$
giving
\be
\label{equ:symmvack}
G^B(k)=\frac{k}{2}.
\ee
The coordinate space correlator is given by the Fourier transform of this, and has a power-law form
\be
\label{equ:symmvac}
G^B(r)=-\frac{1}{2\pi^2r^4}.
\ee
This means that the quantum vacuum fluctuations of the magnetic field are correlated over infinitely long distances. 

It is important to understand that the non-zero space-like two-point function (\ref{equ:symmvac}) does not indicate superluminal signal propagation and violate causality. It describes how the field fluctuations are correlated, not how they propagate.
In cosmology, one would normally argue that nothing can be correlated over more than the particle horizon. However, the way inflation solves the horizon problem is precisely by making the particle horizon larger than the currently observable universe. The power-law form (\ref{equ:symmvac}) should therefore be valid at all observable scales.

In the broken phase, the Higgs mechanism makes the field massive and the vacuum two-point function is
\be
\label{equ:brokenvacuum}
G^B(k)=\frac{k^2}{2\sqrt{k^2+m_\gamma^2}},
\ee
where $m_\gamma$ is the photon mass.
The coordinate space correlation function can be obtained by taking the Fourier transform of this,
\be
\label{equ:vacuumBcorr}
G^B(r)=\frac{m_\gamma^2}{4\pi^2r^2}\left[
\left(m_\gamma r+\frac{2}{m_\gamma r}\right)K_1(m_\gamma r)+K_0(m_\gamma r)
\right],
\ee
where $K_\nu$ is the modified Bessel function of the second kind. The asymptotic long-distance behaviour is exponentially decaying,
\be
\label{equ:brokenvac}
G^B(r)\sim - \frac{2^{1/2}}{8\pi^{3/2}}\frac{m_\gamma^{5/2}}{r^{3/2}}e^{-m_\gamma r}.
\ee

In an adiabatic transition, the correlator would change from Eq.~(\ref{equ:symmvack}) to Eq.~(\ref{equ:brokenvacuum}), and the long-range correlations would disappear. However, as was discussed in Section~\ref{sec:flux}, this cannot happen in a finite time. 
Instead, flux conservation suggests that the longest wavelengths must freeze out and retain their original amplitude (\ref{equ:symmvack}). This is much higher than the amplitude of vacuum fluctuations in the broken phase (\ref{equ:brokenvacuum}), and therefore the transition has turned the quantum vacuum fluctuation into a classical magnetic field.

According to Eq.~(\ref{equ:diffBcorr}), the frozen-out magnetic field correlator after the transition is
\be
\label{equ:frozenBvacuum}
G^B_{\rm final}(k)=\frac{k}{2} e^{-k^2/2\hat{k}^2},
\ee
where the broken phase vacuum contribution (\ref{equ:brokenvacuum}) has been dropped because $m_\gamma$ is generally larger than the scales that are relevant here.
By causality, the critical wavenumber $\vec{k}$ has to be greater than the Hubble rate $|\hat{k}|\gtrsim H$.
The coordinate space correlator is obtained by Fourier transforming this,
\be
G^B_{\rm final}(r)\!=\!\frac{\hat{k}^4}{4\pi^2}\!\left[
\!1\!+\!\sqrt{\frac{\pi}{2}}\left(\frac{1}{r\hat{k}}\!-\!r\hat{k}\right){\rm erfi}\left(\frac{r\hat{k}}{\sqrt{2}}\right)\!e^{-r^2\hat{k}^2/2}\!
\right],
\ee
where $\rm erfi$ is the imaginary error integral. Unsurprisingly, since the long-distance modes are unchanged, the long-distance asymptotic behaviour remains the same as before the transition
\be
\label{equ:vaclongrange}
G^B_{\rm final}(r)\sim -\frac{1}{2\pi^2r^4}.
\ee

According to Eq.~(\ref{equ:stringcorr}), these correlations are imprinted in the string network. Including the contribution from the Kibble-Zurek mechanism~(\ref{equ:KZtaylor}),
Eq.~(\ref{equ:frozenBvacuum}) predicts that the string density correlator should behave at low wavenumbers as
\be
\label{equ:KZFT0pred}
G(k)=\frac{e^2k}{8\pi^2}+\hat\xi k^2+O(k^3).
\ee
Similarly, Eq.~(\ref{equ:vaclongrange}) implies that the asymptotic long-range correlator in coordinate space is
\be
\label{equ:windvaclongrange}
G(r)\sim -\frac{e^2}{8\pi^4 r^4}.
\ee
The strings are therefore correlated even on superhorizon scales, even if the initial state was empty vacuum.

\subsection{Thermal Fluctuations}
\label{sec:thermal}

The correlations are even stronger if the long-wavelength modes of the gauge field were initially not in vacuum. Reheating or any other post-inflationary phenomena cannot create these conditions, because by causality they can only affect sub-horizon modes. However, if inflation lasted only the 50 or 60 e-foldings required to solve the horizon and flatness problems, pre-inflationary field fluctuations would have been imprinted in the cosmic strings. In some models, trans-Planckian physics or phenomena taking place during inflation can also take the gauge field away from its vacuum state.

For concreteness, let us assume that before the start of inflation, sub-horizon magnetic field modes were in thermal equilibrium at temperature $T_{\rm ini}$. During inflation, these comoving modes left the horizon and their temperature was redshifted down to $T=e^{-N}T_{\rm ini},$ where $N\approx 50$ is the number of e-foldings. Today, the wavelength of these modes is comparable to the horizon size, and they are therefore be in the observable range.

At the time of string formation, at the end of inflation, the relevant long-wavelength field modes were therefore in thermal equilibrium at temperature $T$. The equal-time finite-temperature gauge field correlator in the Feynman gauge is
\be
\langle A_i(\vec{k}) A_j(\vec{k}')\rangle
=\delta_{ij}(2\pi)^3\delta(\vec{k}+\vec{k}')\frac{1}{2k}\coth\frac{k}{2T}.
\ee
This gives the magnetic field correlator in momentum space,
\be
\label{equ:GBcoth}
G^B(k)=\frac{k}{2}\coth\frac{k}{2T},
\ee
and, through a Fourier transform, in coordinate space,
\begin{eqnarray}
\label{equ:fullBcorr}
G_{ij}^B(\vec{x})&=&
\left(\delta_{ij}-3\frac{x_ix_j}{r^2}\right)\times\nonumber\\
&&~\times\frac{T^2}{4r^2}\left(1-\coth^2\pi Tr-\frac{\coth\pi Tr}{\pi Tr}\right)\nonumber\\
&&+\left(\delta_{ij}-\frac{x_ix_j}{r^2}\right)\times\nonumber\\
&&~\times\frac{\pi T^3}{2r}\coth\pi Tr\left(1-\coth^2\pi Tr\right),
\end{eqnarray}
where $r=|\vec{x}|$.
The long-distance behaviour is obtained by taking $r\rightarrow\infty$,
\be
\label{equ:thermlongrange}
G_{ij}^B(\vec{x})\rightarrow \left(\delta_{ij}-3\frac{x_ix_j}{r^2}\right)\left(-\frac{T}{4\pi r^3}\right).
\ee

Taking the low wavenumber limit of Eq.~(\ref{equ:GBcoth}) and using Eq.~(\ref{equ:stringcorr}), one find the prediction for the string density correlator,
\be
\label{equ:KZFTpred}
G(k)\approx \frac{e^2T}{4\pi^2}+\hat\xi k^2+O(k^4).
\ee
In coordinate space, Eq.~(\ref{equ:thermlongrange}) implies that the long-distance correlator is
\be
\label{equ:windthermlongrange}
G_{ij}(\vec{x})\rightarrow \left(\delta_{ij}-3\frac{x_ix_j}{r^2}\right)\left(-\frac{e^2T}{16\pi^3 r^3}\right).
\ee
This is traceless, indicating that $G(r)=G_{ii}/2$ decays exponentially and is therefore not a particularly useful quantity for discussing the correlations. Nevertheless, even the trace behaves very differently from the Kibble-Zurek case because winding number is not screened,
\be
\label{equ:thermallowk}
\int d^3x G(\vec{x})=\lim_{k\rightarrow 0} G(k)=\frac{e^2T}{4\pi^2}\ne 0,
\ee
which should be compared with Eq.~(\ref{equ:KZcons}).

Comparison of Eq.~(\ref{equ:windthermlongrange}) with Eq.~(\ref{equ:windvaclongrange}) shows that the correlator decays more slowly than at zero temperature. Because the cosmologically relevant distance scales are extremely long, this means that even a very low non-zero temperature can have significant effects.

\section{Simulations}
\label{sec:simu}

Predictions of the the Kibble-Zurek mechanism and flux trapping have been tested and confirmed in numerical simulations~\cite{Yates:1998kx,Hindmarsh:2000kd,Stephens:2001fv,Donaire:2004gp,BlancoPillado:2007se}. 
Unfortunately these tests are limited to classical field theory, because there are no good ways of simulating non-equilibrium dynamics in quantum field theories, except in the simplest models~\cite{Zurek:2006dr}. For example, the two-particle irreducible effective action formalism~\cite{Aarts:2001qa}, which is a promising tool for perturbative non-equilibrium processes, fails to incorporate topological defects~\cite{Rajantie:2006gy}. Interestingly, it is possible to test these effects experimentally in condensed matter systems which as superfluids and superconductors
\cite{Ruutu:1995qz,Baeuerle:1996zz,PhysRevLett.81.3703,Carmi:2000zz,PhysRevB.60.7595,Maniv:2003zz,PhysRevLett.90.257001,monaco:054509,Golu}, which are, of course, fully quantum mechanical. So far, the emphasis has been on determining the number of strings, and the tests of spatial correlations have so far been limited to short-distance effects~\cite{Rajantie:2001ps,Ray:2001ug,mukai:061704}.

To test the predictions in Section~\ref{sec:FTcorr}, I carried out a set of classical field theory simulations of the gauge theory~(\ref{equ:lag}). The simulations are perfectly conventional, describing the field evolution of the Abelian Higgs model in expanding space. The only difference from usual cosmic string simulations are the initial conditions, which were calculated from the actual thermal equilibrium or vacuum state of the quantum field theory, rather than imposing the usual ad-hoc random phase initial conditions~\cite{Vachaspati:1984dz}. 

\subsection{Non-Zero Temperature}
The first set of simulations focused on the thermal case discussed in Section~\ref{sec:thermal}.
They were carried out on a $256^3$ lattice with comoving lattice spacing $\delta x=1$. The scalar coupling constant was $\lambda=1$, and for the gauge coupling two different values $e=0.1$ and $e=0.5$ were used.

Initially, the system was prepared in classical thermal equilibrium at temperature $T=0.5$, corresponding to the canonical ensemble $p\propto \exp(-\beta H)$. This was done with a Monte Carlo algorithm, more details of which can be found in Ref.~\cite{Donaire:2005qm}.

\begin{figure}
\includegraphics[width=8cm]{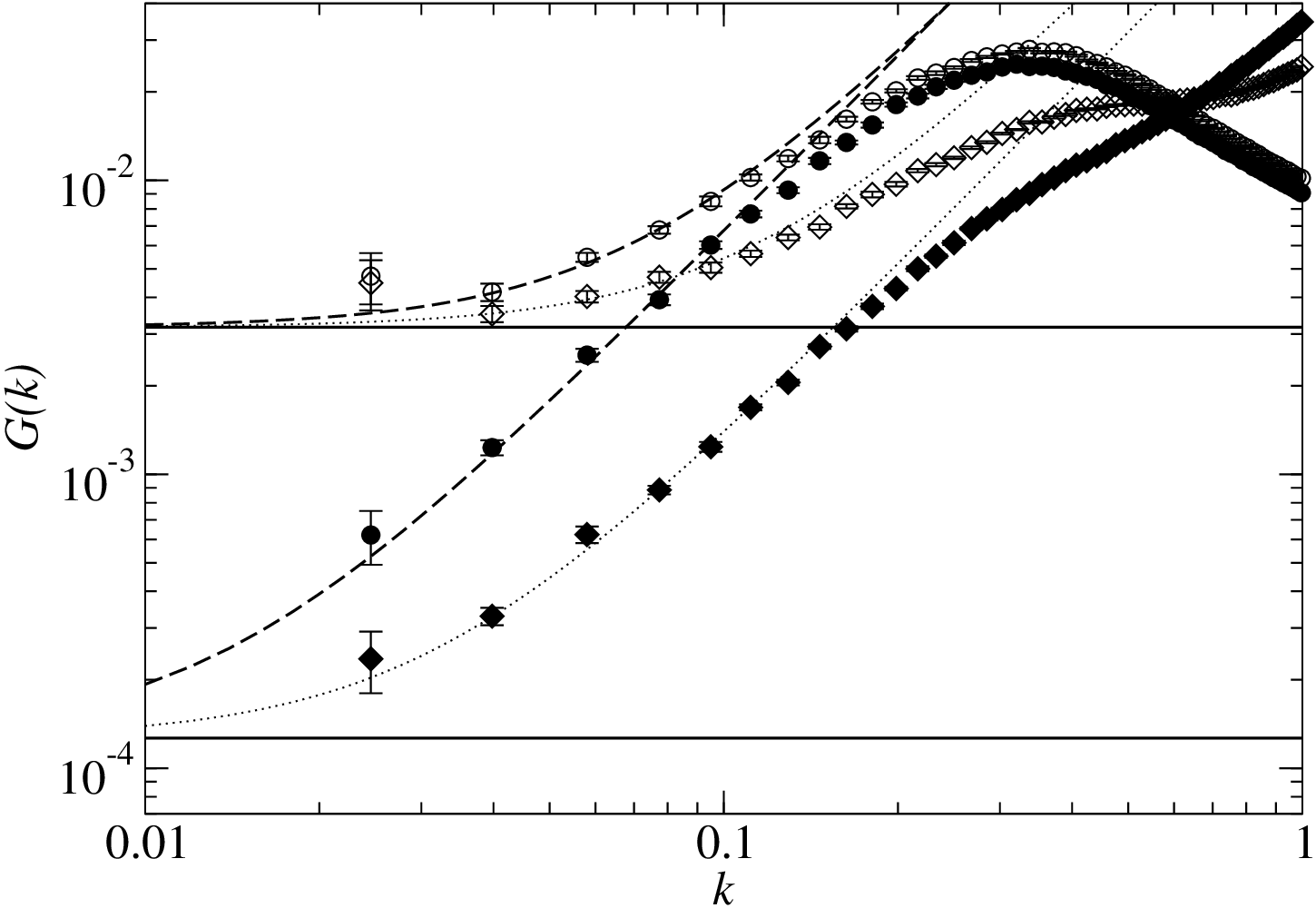}
\caption{\label{fig:twopoint}
The string density correlator $G(k)$ measured at the end of the simulation with thermal initial conditions. The empty and filled diamonds correspond the data for $e=0.5$ and $e=0.1$, respectively, and the circles are the corresponding data after cooling by gradient flow. The two solid lines show the flux trapping prediction (\ref{equ:FTpred}) for $e=0.5$ (upper) and $e=0.1$ (lower). The dashed and dotted lines include an additive contribution $\hat\xi k^2$, as predicted by the Kibble-Zurek mechanism~(\ref{equ:KZFTpred}).}
\end{figure}

The fields were then evolved according to the classical equations of motion. This was done in conformal time $\tau$ defined by $d\tau=dt/a(t)$, where $a$ is the scale factor. In terms of rescaled fields $\tilde\phi=a\phi$ and $\tilde E_i=-\partial_\tau A_i$, the equations of motion are
\begin{eqnarray}
\label{equ:eom}
\partial_\tau^2\tilde\phi&=&
\vec{D}^2\tilde\phi-(m^2a^2-\partial^2_\tau a/a)\tilde\phi-2\lambda|\tilde\phi|^2\tilde\phi,\nonumber\\
\partial_\tau\tilde E_i&=&\partial_j F_{ij}+2e{\rm Im}\tilde\phi^*D_i\tilde\phi,\nonumber\\
\partial_i\tilde{E}_i&=&2e{\rm Im}\tilde\phi^*\partial_\tau\tilde\phi.
\end{eqnarray}
The universe was assumed to be radiation dominated, so that $a=1+H_{\rm ini}\tau$, where $H_{\rm ini}=0.1$ is the initial Hubble rate, and consequently $\partial^2_\tau a/a=0$. The only direct effect of the expansion is therefore the growth of the mass term.
These equations were discretised with constant conformal time step $\delta\tau=0.05$.

To allow a meaningful comparison between the two values of $e$, the mass parameter $m^2$ was set to $m^2=-0.285$ for $e=0.5$ and to $m^2=-0.225$ for $e=0.1$, so that 
in both cases the transition to the broken phase took place near $a\approx \sqrt{2}$.

The local string density was measured at scale factor $a=2$ using the gauge-invariant lattice winding number~\cite{Kajantie:1998bg}, and the two-point correlation function $G(k)$ was calculated. The results are shown as diamonds in Fig.~\ref{fig:twopoint}. 

Because at $a=2$ the temperature is still fairly high, $T=0.25$, thermal fluctuations are present especially at short distance. They should not affect the long-distance behaviour, but to confirm this, the circles in Fig.~\ref{fig:twopoint} show the same data after cooling the system by gradient flow. This removes the fluctuations and makes sure that the strings can really be thought of as classical solutions.

The solid lines show the predicted contribution from flux trapping,
\be
\label{equ:FTpred}
G(k)=\frac{e^2T}{4\pi^2},
\ee
which seems to describe the asymptotic behaviour correctly.

The dashed and dotted lines show the predicted long-distance behaviour (\ref{equ:KZFTpred}),
with Kibble-Zurek correlation length $\hat\xi$ determined as the best fit to the data at $k<0.1$. Without cooling, its value was
\begin{equation}
\hat\xi=
	\begin{cases}
	0.226\pm0.016, & \mbox{for $e=0.5$},\cr
	0.127\pm0.003, & \mbox{for $e=0.1$},
	\end{cases}
\end{equation}
and after gradient flow
\begin{equation}
\hat\xi=
	\begin{cases}
	0.61\pm0.01, & \mbox{for $e=0.5$},\cr
	0.66\pm0.01, & \mbox{for $e=0.1$}.
	\end{cases}
\end{equation}
As one would expect, the correlation length grows during cooling, as the field becomes more ordered, but the long-distance behaviour is unchanged. Note that having $\hat\xi$ less than one does not pose a problem for the lattice discretisation, because $\hat\xi$ is not the actual correlation length but only a fit parameter that is proportional to it. 

The agreement of these predictions with the data at long distances (low $k$) is remarkable. This supports the calculations in Section~\ref{sec:FTcorr}, and suggests that the contributions from the two mechanisms are additive to a good approximation. One can therefore consider the two mechanisms to be essentially independent of each other.

In Fig.~\ref{fig:twopoint} there appears to be a discrepancy between the prediction (\ref{equ:KZFTpred}) and the data at short distances, as the measured correlator turns down at high $k$. However, this does not indicate any conflict between the theory and the data because Eq.~(\ref{equ:KZFTpred}) is only the asymptotic long-range form, and a drop at high $k$ is to be expected.

\subsection{Vacuum}
The second set of simulations tested the zero-temperature case discussed in Section~\ref{sec:vacuum}. 
Following an approach that is widely used in studies of inflationary preheating~\cite{Khlebnikov:1996mc,Prokopec:1996rr,Rajantie:2000fd}, quantum fluctuations were modelled with an Gaussian ensemble of classical fluctuations with the same two-point function. This should be a reasonable approximation to model quantum dynamics by classical fluctuations with the same initial two-point correlation function~\cite{Khlebnikov:1996mc}: At linear level, which should be a reasonably good approximation at early times, the classical and quantum dynamics are identical, and at late times, the occupation numbers of the relevant modes are high and therefore classical approximation should again be valid. 

\begin{figure}
\flushleft
(a)\\
\includegraphics[width=8cm,clip]{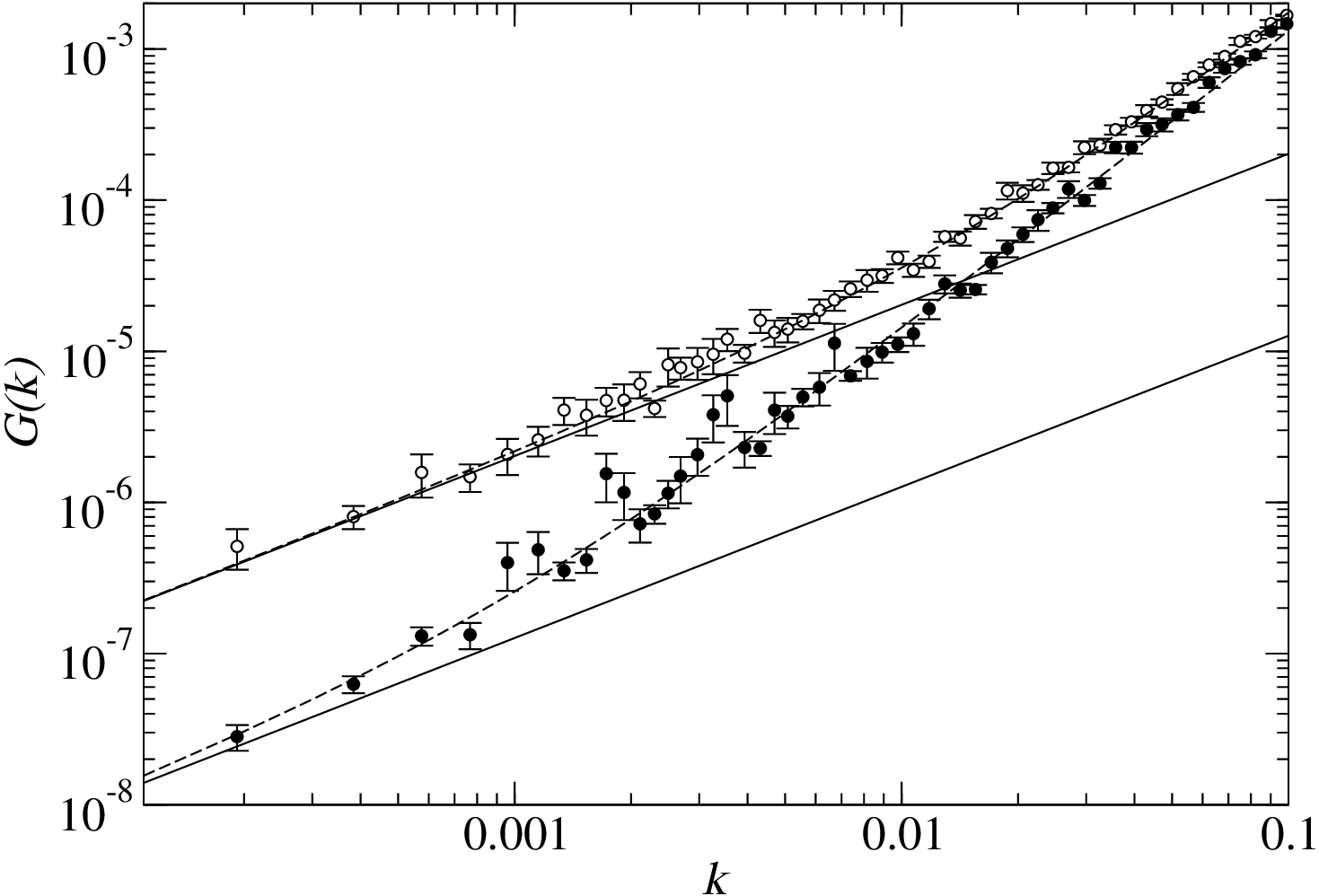}\\[5mm]
(b)\\
\includegraphics[width=8cm,clip]{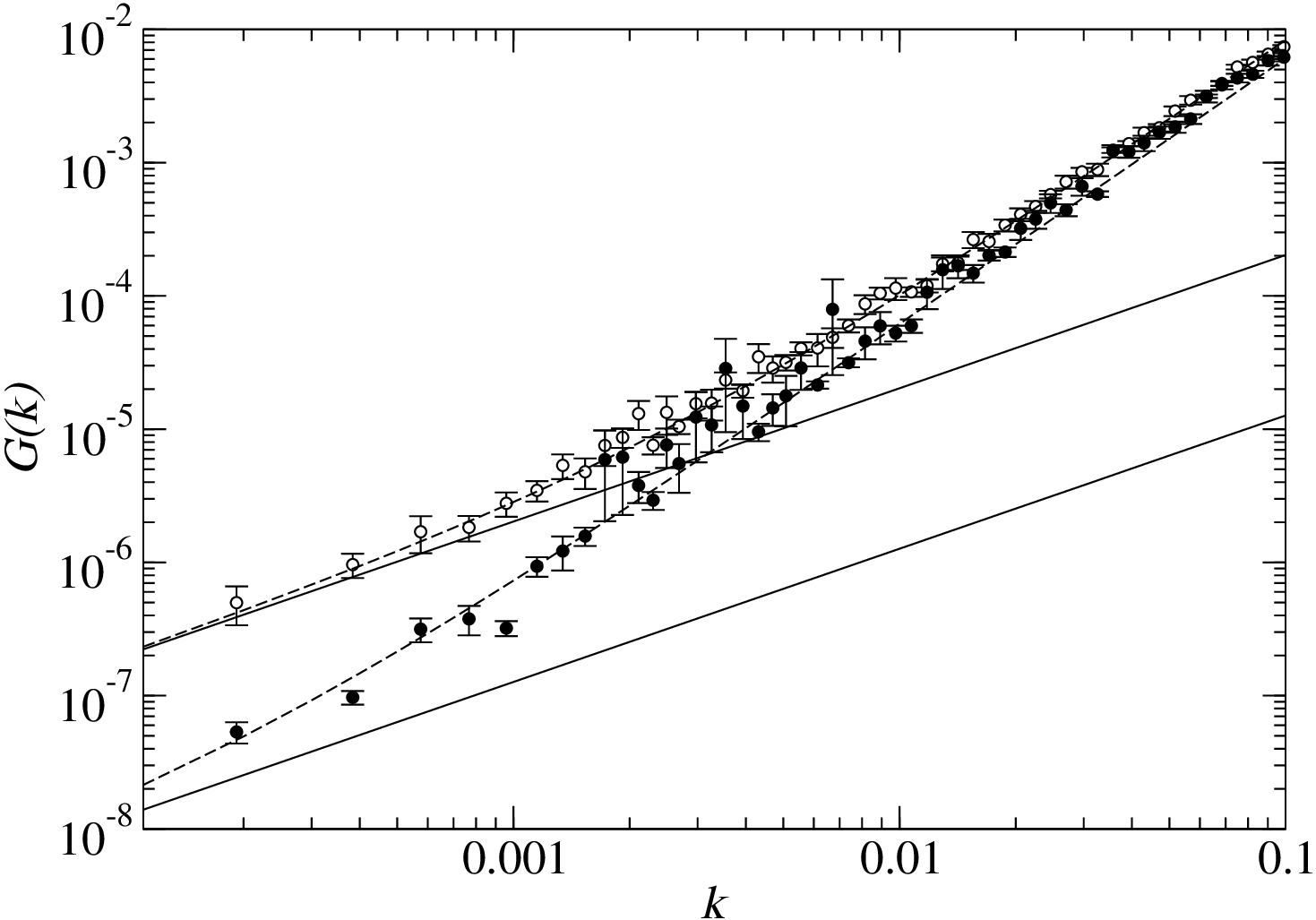}
\caption{\label{fig:vactwopoint}
The string density correlator $G(k)$ measured with vacuum initial conditions (a) at the end of the quench and (b) after gradient flow cooling. The empty and filled circles correspond the data for $e=0.4$ and $e=0.1$, respectively. The two solid lines show the asymptotic behaviour predicted by flux trapping (\ref{equ:FT0pred}) for $e=0.4$ (upper) and $e=0.1$ (lower). The two dashed lines include an additive contribution $\hat\xi k^2$, as predicted by the Kibble-Zurek mechanism~(\ref{equ:KZFT0pred}).}
\end{figure}

Because the predicted correlations are weaker in the vacuum case than at non-zero temperature, larger lattices are needed. To save memory and computing power, and elongated lattice of size $128\times128\times 32768$ was used and the correlations were measured in the long direction. Because even the shorter directions are longer than the light-crossing time, the shape of the lattice should not affect the measured correlations in any way.

Again, the field evolution is given by Eq.~(\ref{equ:eom}), but because the initial state is in vacuum, the expansion alone would not cause a transition in the full quantum theory. It is therefore necessary to have some mechanism that changes the effective mass term, such as coupling to the inflaton field in hybrid inflation. 
For simplicity, the scale factor was set to a constant, $a=1$, and the mass term had linear time dependence 
\be
m^2(\tau)=m^2_{\rm ini}-c\tau.
\ee 
The mass term was initially set to $m^2_{\rm ini}=1$, and it was quenched at rate $c=0.2$ for time $\Delta\tau=10$, so that the final mass term was $m^2_{\rm final}=-1$. 

The string density correlator measured immediately after this quench is shown in Fig.~\ref{fig:vactwopoint}(a), and the same measurement after gradient flow cooling in Fig.~\ref{fig:vactwopoint}(b). Because of the anisotropic lattice shape, the correlator was only measured in the long ($z$) direction, i.e., the measured quantity was
\begin{equation}
G(k)=\frac{1}{2}\left(G_{xx}(k\hat z)+G_{yy}(k\hat z)\right).
\end{equation}

The solid lines show the asymptotic prediction from flux trapping only,
\begin{equation}
\label{equ:FT0pred}
G(k)=\left(\frac{e}{2\pi}\right)^2\frac{k}{2},
\end{equation}
and again they agree with the measured asymptotic behaviour.

The dashed lines show the predicted long-wavelength correlator (\ref{equ:KZFT0pred}), which includes contributions from both flux trapping and the Kibble-Zurek mechanism.
Immediately after the quench [Fig.~\ref{fig:vactwopoint}(a)], the best fit values of the Kibble-Zurek fit parameter $\hat\xi$ were
\begin{equation}
\hat\xi=
	\begin{cases}
	0.154\pm0.002, & \mbox{for $e=0.4$},\cr
	0.131\pm0.003, & \mbox{for $e=0.1$},
	\end{cases}
\end{equation}
and after gradient flow [Fig.~\ref{fig:vactwopoint}(b)],
\begin{equation}
\hat\xi=
	\begin{cases}
	0.812\pm0.010, & \mbox{for $e=0.4$},\cr
	0.605\pm0.020, & \mbox{for $e=0.1$}.
	\end{cases}
\end{equation}
As in the thermal case, the Kibble-Zurek correlation length grows during gradient flow as the system gets more ordered and the smallest string loops disappear, but the long-range asymptotic behaviour remains unchanged and is correctly predicted by the flux trapping mechanism.

\section{Cosmology}
\label{sec:cosmo}

Most calculations of cosmological effects of cosmic strings rely on scaling. The strings were formed in the very early universe and one needs to know their whole cosmological evolution. Because of the vast time and length scales involved, it is impossible to calculate this in detail. However, it is believed that the statistical properties of the string network scale in such a way that, relative to the horizon size, they look the same at all times~\cite{Kibble:1984hp}.

By considering the string density correlator $G(k)$, one can see that this is natural if the strings were formed by the Kibble-Zurek mechanism. At low $k$, the correlator grows as $k^2$, and at some scale it turns down and falls as $1/k$. If we label the position of the maximum by $k_{\rm max}$, we can approximate the correlator by
\begin{equation}
G(k)\approx
	\begin{cases}
	\hat\xi k^2, & \mbox{for}~ k\lesssim k_{\rm max},\cr
	\hat\xi k_{\rm max}^3/k, & \mbox{for}~ k\gtrsim k_{\rm max}.
	\end{cases}
\end{equation}
Expressing this in units of $H$, one has
\begin{equation}
\label{equ:GHubble}
G(\kappa)/H\approx
	\begin{cases}
	(H\hat\xi) \kappa^2, & \mbox{for}~ \kappa\lesssim k_{\rm max}/H,\cr
	(H\hat\xi) (k_{\rm max}/H)^3/\kappa, & \mbox{for}~ \kappa\gtrsim k_{\rm max}/H,
	\end{cases}
\end{equation}
where $\kappa=k/H$.

Because of causality, the correlation length $\hat\xi$ cannot grow faster than the horizon size $1/H$. Assuming that it saturates this bound, the first coefficient $H\hat\xi$ is constant. 

As the string network evolves, its loses energy by radiating it into other degrees of freedom. Because of this, string loops shrink and eventually annihilate. This suppressed the string density correlator, but because this process can only take place on subhorizon scales, it only effects high wave numbers $k\gtrsim H$. Again, if one assumes that this energy loss is efficient so that no power accumulates at high $k$, one concludes that the peak of the correlator moves as $k_{\rm max}\propto H$. In Hubble units, the correlator (\ref{equ:GHubble}) would therefore remain unchanged, which means that the network scales.

The introduction of the thermal flux trapping contribution clearly violates this scaling. Eq.~(\ref{equ:GHubble}) becomes
\begin{equation}
\label{equ:GHubbleT}
G(\kappa)/H\approx
	\begin{cases}
	\frac{e^2T_{\rm s}a_{\rm s}}{4\pi^2 aH}+(H\hat\xi) \kappa^2, & \mbox{for}~ \kappa\lesssim k_{\rm max}/H,\cr
	(H\hat\xi) (k_{\rm max}/H)^3/\kappa, & \mbox{for}~ \kappa\gtrsim k_{\rm max}/H,
	\end{cases}
\end{equation}
where $T_{\rm s}$ and $a_{\rm s}$ are the temperature and the scale factor at the time the strings formed. The extra term grows as $\propto 1/aH$ as the universe expands, ruling out a scaling solution. Eventually, when $aH\sim e^2T_{\rm s}a_{\rm s}$, it starts to have a significant impact even on subhorizon scales.

On the other hand, zero-temperature quantum vacuum fluctuations are still compatible with scaling. When their contribution is included, Eq.~(\ref{equ:GHubble}) becomes
\begin{equation}
\label{equ:GHubblevac}
G(\kappa)/H\approx
	\begin{cases}
	\frac{e^2}{8\pi^2}\kappa+(H\hat\xi) \kappa^2, & \mbox{for}~ \kappa\lesssim k_{\rm max}/H,\cr
	(H\hat\xi) (k_{\rm max}/H)^3/\kappa, & \mbox{for}~ \kappa\gtrsim k_{\rm max}/H.
	\end{cases}
\end{equation}
where the coefficient of the extra term is automatically constant. This suggests that the vacuum fluctuations do not rule out scaling but modify the scaling solution. While this does not exclude the possibility of observable effects, it means that they would necessarily be quite subtle.

The effects of the scaling violation (\ref{equ:GHubbleT}) in the thermal case can be seen by considering the number of ``infinite'' strings. In the cosmic string terminology, this means strings that extend through the whole Hubble volume. The number of infinite strings at any time can be estimated by calculating the average string density in the Hubble volume,
\be
\label{equ:rhoavg}
\vec{\rho}_{\rm avg}\approx \frac{3}{4\pi R^3}\int_{|\vec{x}|<R} d^3x \vec{\rho}(x),
\ee
where $R=1/H$ is the horizon size. Its typical value is given by the square root of the variance,
\bea
\rho_{\rm rms}&=& \sqrt{\langle\vec{\rho}^2_{\rm avg}\rangle}
\nonumber\\
&\approx&\frac{3}{4\pi R^3}\left(\int_{|\vec{x}|,|\vec{y}|<R}
d^3x\,d^3y\,2G(\vec{x}-\vec{y})\right)^{1/2}
\nonumber\\
&=&
\frac{3}{\pi R^3}\left(
\int_0^\infty \frac{dk}{k^4}(kR\cos kR-\sin kR)^2G(k)
\right)^{1/2}
\nonumber\\
&\approx&
\frac{1}{\pi}\left(
\int_0^{1/R} dk\,k^2G(k)
\right)^{1/2},
\label{equ:rhorms0}
\eea
where the contribution from $k>1/R=H$ has been ignored because it is the same as from the Kibble-Zurek mechanism and therefore small.
Substituting Eq.~(\ref{equ:FTpred}) and ignoring numerical factors of order one, one finds that the average string number density in the observable universe is typically
\be
\label{equ:rhorms}
\rho_{\rm rms}\sim \sqrt{\frac{e^2T_{\rm s}a_{\rm s}H^3}{a}}.
\ee
This is due to strings that cross the horizon,
because the net contribution from any loop that is fully inside the observable universe is zero. It be possible to construct a distribution of small loops that would reproduce Eq.~(\ref{equ:rhorms}) as a surface contribution, but Eq.~(\ref{equ:rhorms0}) indicates that it arises from long-wavelength modes with $k<1/R$. The most natural interpretation is therefore that it is given by infinite strings that extend through the whole observable universe.
The typical  number of infinite strings $N_{\rm rms}$ is then obtained by multiplying the density by the cross-sectional area $\pi R^2\sim 1/H^2$ of the observable universe,
\be
N_{\rm rms}\sim \rho_{\rm rms}/H^2\sim \sqrt\frac{e^2T_{\rm s}a_{\rm s}}{aH}.
\ee
As the universe expands, this number grows. There would, therefore, be more and more infinite strings.

Today, with $a=1$ and $H=H_0$, the number of infinite strings would be
\be
\label{equ:N0temp}
N_0\sim \sqrt{\frac{e^2T_{\rm s}a_{\rm s}}{H_0}}.
\ee
Had the relevant long-distance gauge field modes been in thermal equilibrium at the time of string formation, their temperature would have been
$T_{\rm s}=T_c\approx g_*^{-1/3}T_{\rm CMB}/a_{\rm s}$, where $g_*\sim 100$ is the number of relativistic degrees of freedom at that time and $T_{\rm CMB}\approx 2.7\rm K$ is the temperature of the cosmic microwave background radiation. In this case the number of infinite strings today would be huge, $N_0\sim (T_{\rm CMB}/H_0)^{1/2}\approx 10^{14}$, in massive conflict with observations.

In inflationary cosmology, the long-wavelength superhorizon modes would not have been in thermal equilibrium at the time of string formation, as was discussed in Section~\ref{sec:thermal}. Instead, their temperature would be $T_{\rm s}=e^{-N}T_{\rm ini}$, where $N$ is the number of e-foldings and $T_{\rm ini}$ is the pre-inflationary temperature. Therefore we can write Eq.~(\ref{equ:N0temp}) as
\be
N_0\sim \sqrt{\frac{e^2T_{\rm ini}a_{\rm ini}}{H_0}},
\ee
where $a_{\rm ini}=e^{-N}a_{\rm s}$ is the scale factor at the start of inflation. In order for inflation to solve the horizon problem, the comoving horizon size at the start of inflation must have been larger than today,
i.e., $H_{\rm ini}a_{\rm ini}<H_0$.
If this bound is saturated, one finds
\be
N_0\approx\sqrt{\frac{e^2T_{\rm ini}}{H_{\rm ini}}}.
\ee
In order to have inflation, the temperature $T_{\rm ini}$ can not have been so high that is would have dominated the energy density. This gives a bound
\be
H^2_{\rm ini}\gtrsim \frac{g_*T_{\rm ini}^4}{M_{\rm Pl}^2},
\ee
where $g_*$ is the number of degrees of freedom and $M_{\rm Pl}=(8\pi G)^{-1/2}$ is the reduced Planck mass.
This bound is saturated if the temperature was initially even higher and inflation started when radiation became subdominant. In this case, one finds
\be
N_0\approx\sqrt{\frac{e^2M_{\rm Pl}}{g_*^{1/2}T_{\rm ini}}}.
\ee
This can be as high as a dozen for GUT scale inflation, and in low-energy inflationary models it can be much higher.

The presence of these infinite strings may affect on the evolutions of the string network and have indirect observable consequences. Their contribution to the total energy density would be
\be
\Omega_{\rm inf}\approx \frac{\mu N_0}{3M_{\rm Pl}^2},
\ee
which can be significant if $N_0$ is large. Furthermore, they would single out a preferred direction, that of $\vec{\rho}_{\rm avg}$ in Eq.~(\ref{equ:rhoavg}), thereby breaking the isotropy of the universe.
It may also be possible to observe them directly, and count them, using the Kaiser-Stebbins effect~\cite{Kaiser:1984iv}, which creates discontinuities in the temperature of the cosmic background. 

\section{Conclusions}
The results in this paper show that gauged cosmic string formation is dominated by trapping of gauge ``magnetic'' flux at long distances, and the gauge field configuration becomes imprinted in the string network. Because of the preceding period of inflation and the masslessness of the gauge field, this can give rise to correlations over massively superhorizon scales at the time of string formation and can therefore have potentially observable consequences.

The detailed form of the string correlations was calculated for thermal and vacuum initial states, corresponding to different cosmological scenarios. The correlations are stronger in the former case, but they are also present in the latter, cosmologically more realistic case, which describes a phase transition following a long period of inflation. 
In that case, these correlations are not generated by inflation, and that the findings are therefore not in conflict with the well-known result~\cite{Turner:1987bw} that inflation does not generate long-range magnetic fields. Instead, the correlations are a property of the vacuum quantum fluctuations, which are turned into physical string excitations by the phase transition. The role of inflation is simply to remove all other fluctuations, thermal or otherwise, which would lead to stronger correlations. The vacuum result is therefore not specific to inflationary cosmology, and  be seen as a general lower bound given by the uncertainty principle.

The results were confirmed by classical lattice field theory simulations. Although this is a widely used approach, its validity in the zero-temperature case is uncertain, because the whole effect originates in vacuum fluctuations are is therefore fully quantum mechanical. Interestingly, the predictions can also be tested in superconductor experiments~\cite{Golu}, which are obviously fully quantum mechanical and therefore avoid this potential pitfall completely.  

The calculations in this paper were carried out in the Abelian Higgs model, as is the case with essentially all studies of cosmic strings. The qualitatively new behaviour is therefore not a consequence of adding new degrees of freedom or modifying the microscopic theory in other ways. Instead, it arises because instead of imposing {\it ad hoc} initial conditions, the fields are set to be in their thermal equilibrium or quantum vacuum state. Because the arguments used were based on general physical principles, the results should also generalise to essentially any model of local cosmic strings that carry a gauge flux, including cosmic superstrings formed in brane collisions. 

The observational effects of these correlations are not yet known. It was shown that sufficiently strong thermal fluctuations will violate the usual scaling assumption, and even at zero temperature, the precise scaling solution would have to change. This may well affect the cosmic string contribution to the CMB power spectrum. There may also be totally new effects. For instance, one can speculate whether the preferred direction of the infinite strings produced by flux trapping could be observed. Also, the difference between string networks in different Hubble volumes are not random, as in the conventional pictures, but correlated over infinite distances. Since they influence the expansions of the universe locally in each Hubble volume, they can give rise to potentially observable density perturbations. Determining these signatures will require large-scale numerical simulations of the string network evolution. A key lesson from this paper is that the usual random phase initial conditions, which are a concrete realisation of the Kibble-Zurek mechanism, are insufficient and, instead, full quantum or thermal initial conditions should be used. 

\section*{Acknowledgments}
The author would like to thank N.~Bevis for useful discussions and STFC for financial support. 
This research was carried out on the COSMOS supercomputer (SGI Altix 4700) funded by STFC, HEFCE and SGI.

\bibliography{stringcorr_rev}{}
\bibliographystyle{apsrev}

\end{document}